# On the Universality of Mesoscience: *Science of 'the in-between'*

Jinghai Li*[1], Wenlai Huang[1], Peter P. Edwards*[2], Mooson Kwauk[1],
John T. Houghton[3] & Daniel Slocombe[2]

[1]State Key Laboratory of Multiphase Complex Systems, Institute of Process Engineering, Chinese Academy of Sciences, Beijing 100190, P. R. China.

[2]Department of Chemistry, University of Oxford, Oxford, OX1 3QR, UK.

[3]Hadley Centre, Meteorological Office, Exeter, EX1 3PB, UK.

*To whom correspondence should be addressed. E-mail: jhli@home.ipe.ac.cn (J. Li); peter.edwards@chem.ox.ac.uk (P. P. Edwards)

**The universality of mesoscales, ranging between elemental particles and the universe, is discussed here by reviewing widely disparate fields and presenting four cases, at differing hierarchical levels, from chemistry, chemical engineering, meteorology, through to astronomy. An underpinning concept, "*Compromise in competition*", is highlighted between various dominant, but competing mechanisms, and is identified here to be the universal origin of complexity and diversity in such examples. We therefore advance this as a key underlying principle of an emerging science — Mesoscience.**

**One Sentence Summary:** Compromise in, and between, competing mechanisms is identified as the universal origin of complexity and diversity, and forms the core of Mesoscience.

**"Meso" is being highlighted world-wide, but is this a singular "mesoscale" or the plural "mesoscales"?** While the U.S. Department of Energy (DOE) is currently exploring plans to launch a major new research initiative in mesoscale science[1], the National Natural Science Foundation of China (NSFC) recently launched a research program in process engineering entitled *"Mechanism and manipulation at mesoscales in multiphase reaction processes"* (http://www.nsfc.gov.cn/Portal0/InfoModule_584/50112.htm). The former initiative centres upon the nature and importance of mesoscale science occurring between the quantum world of atoms and atomic clusters and that of macroscopic-scale bulk materials. The latter initiative identifies two naturally-related mesoscales relating to interfacial phenomena and the heterogeneity of flow structures in functioning chemical reactors. We pose the question here: "*Should this emerging area be advanced along the lines of the DOE's 'atom-to-bulk' mesoscale science or, the NSFC's science and engineering for two mesoscales, or, should it be perhaps be even broader, encompassing the full diversity of natural phenomena for mesoscales across the entire spectrum of science and technology?*" This latter perspective is based on the possibility that there may be a common, governing, principle for the science of mesoscales, even though they show a remarkable diversity of phenomena[2,3].

We believe that an exciting challenge, therefore, is now to recognize that the field of *"Mesoscience"* should be the science of *ALL* occurring mesoscales which exist between elementary particles and the universe. Thus, *"Meso"* should not simply be viewed in terms of the sole, unitary physical dimension of characteristic size but, rather, identified as a unifying concept signifying the *"intermediate"* or *"in-between"* regimes, which straddle the domain of complex systems from the *"small scale"* of individual elements, to the *"large scale"* of collective systems or ensembles[2]. We attempt here to highlight and comment upon some of the significant issues in this important, emerging field.

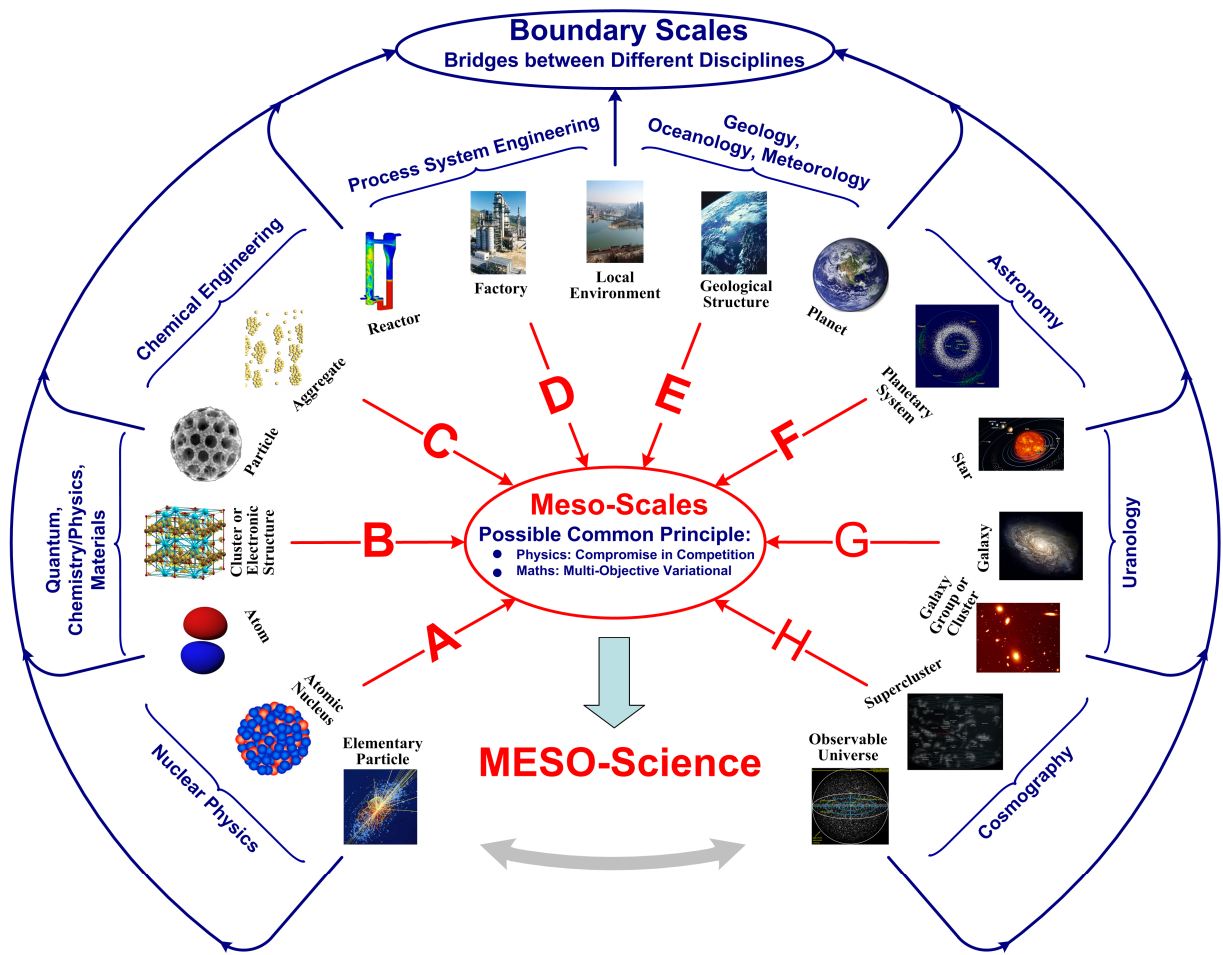

**Figure** | *A unified theory of Mesocience will encompass all mesoscale phenomena: This represents a common challenge for the whole spectrum of science and technology.*

**Our proposal here is that the concept of "Compromise in competition" is likely to be a unifying principle,** governing all mesoscales, providing the basis for, and rationale of Mesoscience as the centrally important science of 'the in-between'. Mesoscale phenomena exist all around us. There are different branches (or bifurcations) bridging elementary particles through to the observable universe, all of which feature the multi-level and multi-scale characteristics of the natural world. Each such branch is of-itself multi-leveled, and each level consists of a "small scale" and a "large scale", bridged by the ubiquitous "in-between" mesoscale. Thus, elements for each level form systems, which are, themselves, then the elements of even larger systems at a higher level, and so on. Such structures and phenomena pervade the entire spectra of space and of time, covering length, energy and time scales from elementary particles to the observable universe[2-8], as shown in the Figure. Mesoscale thus refers to intermediate or "in-between" phenomena, mechanisms and processes between "small" elements and "large" systems[2]. The exploration of IPE (Institute of Process Engineering, Chinese Academy of Sciences) since the 1980's in the fields of chemical and process engineering has identified a key concept which governs the understanding and rationalization of such mesoscales; this concept is *"Compromise, in competition, between different dominant mechanisms"*. That is, dominant mechanisms and processes, each formulated as an extremum, of necessity imbue compromise to any intermediate, or "in-between" system, then to yield an overall steady *"meso-state"*[3,9,10]. We have advanced the view that compromise arising out of such competition is the origin of diverse mesoscale phenomena and physico-chemical properties in process and chemical engineering. Mathematically, this is formally akin to the so-called multi-objective variational problem[10], which, we believe, is most challenging and deserves great attention due to its universal importance. Although this common principle of compromise in competition needs to be further confirmed in physics, and formulated in detail in mathematics, its universality was indeed anticipated in earlier work[9,11] and explored subsequently in many different problems[12,13]. Thus, one finds:

- ***In materials chemistry and physics:*** Towards the *lower level hierarchy* (mesoscale B in the spectrum of the Figure), the rules of compromise are highly evident. To take one example; in venerable colloidal systems, we see compromise arising from competition between natural forces, which defines not only the structures, but also the physico-chemical properties of colloids[14]. "Mesoscopic" colloids are classically defined as systems involving particles ranging from 10 to 1,000 nm in diameter, appearing as clusters or aggregates, usually with complex structures. Such structures are synthesized (through nucleation and growth) and maintained (stabilization) through compromise between the phenomena of *association and dispersion*, and driven by competing, natural forces. If the former dominates, flocculation occurs, and once the latter dominates, a homogeneous solution results[15]. Thereby, colloidal systems are stable only if the two competing trends compromise. Upon any change in the controlling parameters, e.g., the ion strength, the temperature, the pH value, etc, colloidal systems can be destabilized since either single trend alone becomes dominant. Many examples at this mesoscale level also follow the principle of compromise in competition. For example, in condensed matter physics, the Mott Metal-Insulator Transition (MIT) describes the electronic phase transition from the metallic, conducting state to the insulating, non-conducting state in solids and liquids. For the Mott Transition, competition is found between the Coulomb repulsive attraction between electrons, which causes electrons to become localized, and screening (reflected in the electronic bandwidth) which favours delocalization of charge carriers[16]. The location of the MIT is captured in the universal Mott criterion, which is a function of electron density and Bohr radius of the isolated (localized) centers[17]. This remarkable analogy with compromise in competition has also been recognized[18] with the description of "electronic mayonnaise" or phase separation.

- ***In chemical or process engineering:*** At the *middle level hierarchy* of physical science (see mesoscale C in the Figure), for instance, in gas/solid-particle flow, with increasing gas velocity, the system can show three different regimes with distinct characteristics (particle-dominating, particle-gas compromising through to gas-dominating). At low gas velocity, below a minimum fluidization velocity, solids dominate the ensuing structure, featuring minimum packing voidage; At very high gas velocity, gas dominates the solids, showing a uniform, very dilute state, characterized by minimal energy consumption for suspending and transporting particles per unit volume[9]; In the middle range of gas velocity, the gas-solid compromising regime shows up, that is, at any time and at any spatial position, either solid particles dominate to realize minimal packing voidage or gas dominates to realize minimal energy consumption for suspending and transporting particles, but, absolutely and importantly, not both. Gas and particles achieve the necessary level of compromise with each other in achieving their respective, dominant roles, that is, alternately both in time and in space[9]. It is evident that when either the gas or the solid dominates the other, the system tends to be uniform and simple. Only when neither can dominate the other, both gas and particles have to compromise with each other to coexist, leading to complexity and heterogeneity, that is, a complex phenomenon or mechanism originates from the compromise between different simple phenomena and mechanisms. Without considering this kind of compromise between gas and solids at mesoscale of computational grids, computational fluid dynamics would give unacceptable errors[9].

- *In meteorological and climate science:* At the *upper level hierarchy* (mesoscale E of the Figure), observations and computer simulations are applied to provide forecasts of future weather or climate over the whole globe – something of very great economic, social and political importance in today's world. Computer simulations (often called models) are essential to add together all the nonlinear processes and flows occurring in the atmosphere or in the whole climate system. A major problem with these simulations is how to include motions at the smallest scales that cannot be explicitly formulated or described. Mostly this has been done by lumping together all motion below the grid size in a simple formula expressed in terms of grid size variables. Such a formulation allows for no interactions between the smallest scale flows and characteristics of the flows at larger scales, that is, the connections discussed above are missing. This is a severe limitation as has been increasingly realized. Detailed observations of flows at different scales identify consistent and structured interactions, correlated between scales, which need to be included in the model formulations. Substantial improvements in the simulations are thereby being realized[19]. This is again a good example of Mesoscience providing a natural bridge between small scale flows (range of metres to kilometres) and large scale flows (range of kilometres to hundreds of kilometres) in the atmosphere, in a manner similar to those described above for gas-solid flows and materials for a very different range of scales and interactions.

- *In astronomy or uranology:* The same fundamental challenges are faced at even larger hierarchical structures (mesoscale G in the Figure); here we encounter the challenge of modeling convective energy transport in the subsurface layers of the sun and other cool stars. A hierarchy of convection cells forms, the largest of which reaching up to 30,000km. Mesogranular formations have been modeled numerically, but there are severe limitations because of the ignorance of smaller-scale structures in coarse granular approaches[8]. Structural interactions, correlated between length-scales in a manner similar to the meteorological example are here identified in the form of mesogranulation. Again, the formation of such mesoscale structures might be owing to the compromise between competing dynamic mechanisms, driven by rotation, divergent motion, etc. At the galactic level, complex structures have also been well recognized, e.g., the Hubble classification exists. Such structures arise at a mesoscale between the stellar scale and the galaxy group or cluster scale, and reflect the compromise among different dynamic mechanisms (e.g., gravitational capture, radiation dispersion, etc), usually expressed in reaction-diffusion models[20].

**Progresses in Mesoscience will hopefully lead to breakthroughs in both science and technology:** In the above four examples we have attempted to illustrate how Mesoscience reflects its important status as a natural bridge – the "in-between" regime – from small-scale to large-scale, representing a more attractive alternative to complexity science in contributing to its development[2,3]. Further *p*rogress must centre upon establishing such bridges between the multifarious constituent "elements" and "systems or ensembles" (see Figure) to pave the way to progress in controlling the science and technology of complete-system properties. This can come from an appreciation and understanding of mesoscale structures and properties[13].

We believe that further progress in Mesoscience will also help catalyze computational science to a higher, *"greener"* level. Understanding mesoscales, particularly when the (common) unifying principle of *"Compromise in competition"* can be established, opens the possibility of optimizing communication, computation, and storage at different scales and between scales, and holds the key to consistency between physical modeling, numerical algorithms and computational hardware structure[21]. Such advances will significantly reduce the gap between the theoretical peak of computer capacity and the real operational capability of computation; this, a current, forefront challenge in computer science and technology.

Although this article discusses only four representative (but broad) branch levels of the physical and engineering sciences, we believe that the concepts introduced here will be applicable and relevant to even broader branches of knowledge, viz., the natural life and social sciences. In fact, these four examples from different levels of the hierarchical spectrum in the Figure show interesting – and important – similarities in understanding such mesoscales for systems where correlation between scales is crucial.

**There are important implications for an emerging field of Mesoscience; the challenge is to attract and capture insights from different, superficially disparate fields:** There have been several important and significant earlier advances in this regard. The broad applicability of Mesoscience across ALL mesoscales has been recognized and advanced in the field of process engineering[2,3,10,12,13]. Self-assembly in chemistry was innovatively recognized to be able to expand naturally from molecules to galaxies with similar, but not identical, rules[6], thus providing the key, unified science of self-assemblies of various sizes of components. We believe that the similarities among those important, guiding rules in self-assembly could indeed be related to the principle of compromise in competition[2,9,11,12]. Intriguingly, in the field of chemical and process engineering, particle clustering in gas-solid flows closely corresponds to the process of dynamic self-organization driven by dissipating energy[22-24].

The challenge of mesoscales in a variety of fields is now attracting increasing attention as the vehicle for *"Bridging the small scale and the large scale"*, across the disciplines of chemistry[6,7,18,25], biology[26], cosmology[27], atmospheric science[28], polymer science[29], through to the social sciences[30]. Perhaps the most active field is that of meteorology and related fields where the term "mesoscale" has been identified and utilized for several decades[31]. Existing theoretical achievements such as dissipative structures[23], synergetics[24], scaling and renormalization theories[32], reflect prominent efforts along this direction.

To correlate and connect the small-scale with large-scale, either within a single level or for a whole spectrum of multiple levels as show in the Figure, is the common theme of complexity science. We believe that the recognition of the universality of the common principle of compromise in competition for all mesoscales will catalyse its development and application.

**Mesoscience — Surely, a big(ger) thing:** In summary, the field of Mesoscience is now a developing, forefront activity in both the scientific and engineering communities. We strongly believe that this, as a broad, emerging science, should encompass cognisance of – or, at least an awareness of – all possible mesoscales across the natural world (see the Figure). Such an advance will benefit from a search for common, underlying and unifying principles (e.g., Compromise in competition). In addition to studying systems across individual disciplines, natural trans-disciplinarity is critical to making Mesoscience *"The next big(ger) thing*[1]*"*.

**References**

1. Service, R. F. The next big(ger) thing. *Science* **335**, 1167 (2012).

2. Li, J., Ge, W. & Kwauk, M. Meso-scale phenomena from compromise — A common challenge, not only for chemical engineering. *arXiv:0912.5407v4*.

3. Li, J., Ge, W., Wang, W. & Yang, N. Focusing on the meso-scales of multi-scale phenomena — In search for a new paradigm in chemical engineering. *Particuology* **8**, 634-639 (2010).

4. Marquadt, P. & Nimtz, G. The size-induced metal-insulator transition and related electron interference phenomena in modern microelectronics. In *Festkörperprobleme* **29**, Rössler, U., Ed. (Vieweg, Braunschweig, 1989), pp. 317-328.

5. Edwards, P. P., Johnston, R. L. & Rao, C. N. R. in *Metal Clusters in Chemistry*, Braunstein, P., Oro, L. A. & Raithby, P. R., Eds. (Wiley-VCH, 2000), chap. 4.8.

6. Whitesides, G. M. & Boncheva, M. Beyond molecules: Self-assembly of mesoscopic and macroscopic components. *Proc. Natl. Acad. Sci.* **99**, 4769-4774 (2002); Whitesides, G. M. & Grzybowski, B. Self-assembly at all scales. *Science* **295**, 2418-2421 (2002).

7. Cölfen, H. & Mann, S. Higher-order organization by mesoscale self-assembly and transformation of hybrid nanostructures. *Angew. Chem. Int. Ed.* **42**, 2350-2365 (2003).

8. Ploner, S. R. O., Solanki, S. K. & Gadun, A. S. Is solar mesogranulation a surface phenomenon? *Astron. Astrophys.* **356**, 1050-1054 (2000).

9. Li, J., Tung, Y. & Kwauk, M. Multiscale modeling and method of energy minimization in particle-fluid two-phase flow. In *Circulating Fluidized Bed Technology II*, Basu, P. & Large, J. F., Eds. (Pergamon, London, 1988), pp. 89-103; Li, J. & Kwauk, M. *Particle-Fluid Two-Phase Flow — The EMMS Method* (Metallurgical Industry Press, Beijing, 1994); Li, J. Multiscale-modeling and method of energy minimization for particle-fluid two-phase flow. *Ph.D. Thesis*, (Institute of Chemical Metallurgy, Chinese Academy of Sciences, Beijing, 1987).

10. Li, J., Zhang, J., Ge, W. & Liu, X. Multi-scale methodology for complex systems. *Chem. Eng. Sci.* **59**, 1687-1700 (2004).

11. Li, J., Qian, G. & Wen, L. Gas-solid fluidization: A typical dissipative structure. *Chem. Eng. Sci.* **51**, 667-669 (1996); Li, J., Cheng, C., Zhang, Z., Yuan, J., Nemet, A. & Fett, F. N. The EMMS model — Its application, development and updated concepts. *Chem. Eng. Sci.* **54**, 5409-5425 (1999).

12. Li, J. & Kwauk, M. Multiscale nature of complex fluid-particle systems. *Ind. Eng. Chem. Res.* **40**, 4227-4237 (2001).

13. Ge, W., Wang, W., Yang, N. *et al.* Meso-scale oriented simulation towards virtual process engineering (VPE) — The EMMS paradigm. *Chem. Eng. Sci.* **66**, 4426-4458 (2011).

14. Weiser, H. B. *A Textbook of Colloid Chemistry* (Wiley, New York, ed. 2, 1939).

15. Prost, J. & Rondelez, F. Structures in colloidal physical chemistry. *Nature (Supplement)* **350**, 11-23 (1991).

16. Mott, N. F. The basis of the electron theory of metals, with special reference to the transition metals. *Proc. Phys. Soc. A* **62**, 416-422 (1949); Metal-insulator transition. *Rev. Mod. Phys.* **40**, 677-683 (1968).

17. Edwards, P. P. & Sienko, M.J Universality aspects of the metal-nonmetal transition in condensed media. *Phys. Rev. B* **17**, 2575-2581 (1978).

18. Schmalian, J. & Wolynes, P. G. Electronic mayonnaise: Uniting the sciences of "hard" and "soft" matter. *MRS Bull.* **30**, 433-436 (2005).

19. Palmer, T. N. Towards the probabilistic Earth-system simulator: A vision for the future of climate and weather prediction. *Q. J. R. Meteorol. Soc.* **138**, 841-861 (2012).

20. Cartin, D. & Khanna, G. Self-regulated model of galactic spiral structure formation. *Phys. Rev. E* **65**, 016120 (2001).

21. Chen, F., Ge, W., Guo, L., He, X., Li, B., Li, J., Li, X., Wang, X. & Yuan, X. Multi-scale HPC system for multi-scale discrete simulation — Development and application of a supercomputer with 1 Petaflops peak performance in single precision. *Particuology* **7**, 332-335 (2009).

22. Ball, P. *The Self-Made Tapestry: Pattern Formation in Nature* (Oxford Univ. Press, 1999).

23. Prigogine, I. & Nicolis, G. *Self-Organization in Non-Equilibrium Systems* (Wiley, 1977).

24. Haken, H. *Synergetics, an Introduction: Nonequilibrium Phase Transitions and Self-Organization in Physics, Chemistry, and Biology* (Springer-Verlag, New York, 1983).

25. Antonietti, M. & Ozin, G. A. Promises and problems of mesoscale materials chemistry or why meso? *Chem. Eur. J.* **10**, 28-41 (2004); Fan, J., Boettcher, S. W., Tsung, C. -K., Shi, Q., Schierhorn, M. & Stucky, G. D. Field-directed and confined molecular assembly of mesostructured materials: Basic principles and new opportunities. *Chem. Mater.* **20**, 909-921 (2008).

26. Walsh, C. Chemical biology: Information, mesoscale science and the engineering ethos. *Chemistry & Biology* **5** (8), R177-R179 (1998); Tozzini, V. Multiscale modeling of proteins. *Acc. Chem. Res.* **43**, 220-230 (2010).

27. Clarkson, C., Ellis, G., Larena, J. & Umeh, O. Does the growth of structure affect our dynamical models of the Universe? The averaging, backreaction, and fitting problems in cosmology. *Rep. Prog. Phys.* **74**, 112901 (2011).

28. Zhang, Y., Bocquet, M., Mallet, V., Seigneur, C. & Baklanov, A. Real-time air quality forecasting, part I: History, techniques, and current status. *Atmos. Environ.* **60**, 632-655 (2012).

29. Ramanathan, M. & Darling, S. B. Mesoscale morphologies in polymer thin films. *Prog. Polym. Sci.* **36**, 793-812 (2011).

30. Brännström, Å. & Sumpter, D. J. T. The role of competition and clustering in population dynamics. *Proc. R. Soc. B* **272**, 2065-2072 (2005); Lehner, E. Describing students of the African Diaspora: Understanding micro and meso level science learning as gateways to standards based discourse. *Cult. Stud. Sci. Edu.* **2**, 441-473 (2007).

31. Fujita, T. Precipitation and cold air production in mesoscale thunderstorm systems. *J. Meteorol.* **16**, 454-466 (1959); Orlanski, I. A rational subdivision of scales for atmospheric processes. *Bull. Am. Meteorol. Soc.* **56**, 527-530 (1975).

32. Wilson, K. G. Problems in physics with many scales of length. *Sci. Am.* **241**, 158-179 (1979).

**Acknowledgements** We greatly appreciate the long-term support from the National Natural Science Foundation of China, and the Einstein Professorship Program of the Chinese Academy of Sciences.

**Author Contributions** All the authors jointly contributed to this manuscript, based on their disciplinary expertise.